\documentclass[twocolumn,prb,floatfix,superscriptaddress,nobibnotes]{revtex4-2}
\usepackage[unicode=true, colorlinks=true, citecolor={blue!80!black}, urlcolor={blue!50!black}, linkcolor = {blue!80!black}]{hyperref}
\usepackage{graphicx}
\usepackage{bm}
\usepackage{array}
\usepackage{amsmath}
\usepackage{amssymb}
\usepackage{contour}

\usepackage{physics}
\usepackage[normalem]{ulem}
\usepackage{xcolor}

\usepackage{nicematrix}

\usepackage{cleveref}

\makeatletter
\let\ORIbbl@fixname\bbl@fixname
\def\bbl@fixname#1{%
  \@ifundefined{languagealias@\expandafter\string#1}
    {\ORIbbl@fixname#1}
    {\edef\languagename{\@nameuse{languagealias@#1}}}%
}
\newcommand{\definelanguagealias}[2]{%
  \@namedef{languagealias@#1}{#2}%
}
\makeatother
\definelanguagealias{en}{english}
\definelanguagealias{EN}{english}

\newcounter{appendixpar}
\renewcommand{\theappendixpar}{Appendix~\Alph{appendixpar}}

\newcommand{\appendixpar}[1]{%
  \refstepcounter{appendixpar}%
  \textit{\theappendixpar}: \textit{#1}.%
}

\newcommand{\cpar}[1]{%
    \textit{#1}.%
}

\begin{document}

\title{Non-Hermitian skin effect and electronic nonlocal  transport}

\date{\today}

\author{Carlos Payá}
\thanks{These authors contributed equally to this work.}
\affiliation{Instituto de Ciencia de Materiales de Madrid (ICMM), CSIC, 28049 Madrid, Spain}
\affiliation{Center for Quantum Devices, Niels Bohr Institute, University of Copenhagen, DK-2100 Copenhagen, Denmark}
\author{Oliver Solow}
\thanks{These authors contributed equally to this work.}
\affiliation{Center for Quantum Devices, Niels Bohr Institute, University of Copenhagen, DK-2100 Copenhagen, Denmark}
\author{Elsa Prada}
\affiliation{Instituto de Ciencia de Materiales de Madrid (ICMM), CSIC, 28049 Madrid, Spain}
\author{Ramón Aguado}
\affiliation{Instituto de Ciencia de Materiales de Madrid (ICMM), CSIC, 28049 Madrid, Spain}
\author{Karsten Flensberg}
\email{flensberg@nbi.dk}
\affiliation{Center for Quantum Devices, Niels Bohr Institute, University of Copenhagen, DK-2100 Copenhagen, Denmark}

\begin{abstract}
    Open quantum systems governed by non-Hermitian \emph{effective} Hamiltonians exhibit unique phenomena, such as the non-Hermitian skin effect, where eigenstates localize at system boundaries. We investigate this effect in a Rashba nanowire coupled to a ferromagnetic lead and demonstrate that it can be detected via non-local transport spectroscopy: while local conductance remains symmetric, the non-local conductance becomes non-reciprocal. We account for this behavior using both conventional transport arguments and the framework of non-Hermitian physics. Furthermore, we explain that exceptional points shift in parameter space when transitioning from periodic to open boundary conditions, a phenomenon observed in other non-Hermitian systems but so far not explained. Our results establish transport spectroscopy as a tool to probe non-Hermitian effects in open electronic systems.
\end{abstract}
\maketitle

\cpar{Introduction}
Recently, there has been an increased interest in the physics of open quantum systems described by non-Hermitian \emph{effective} Hamiltonians \cite{Minganti:PRA19, Minganti:PRA20, Ashida:AP20}. Introducing non-Hermitian terms changes the spectrum of the system, most dramatically through the emergence of exceptional points (EPs) \cite{Heiss:JPAMT12, Bergholtz:RMP21,Cayao:PRB25}, points in parameter space where the Hamiltonian is not diagonalizable. EPs  are closely tied to the topological classification of non-Hermitian Hamiltonians \cite{Kawabata:PRX19}, where the bulk-boundary correspondence breaks down \cite{Yao:PRL18, Xiong:JPC18}. Consequently, topological phase transitions observed under periodic boundary conditions (PBC) cannot be directly related to those occurring in the same system with open boundary conditions (OBC).

Another fascinating aspect of non-Hermitian physics is the non-Hermitian skin effect (NHSE), which involves eigenstate localization at the system boundaries \cite{Ashida:AP20, Zhang:NC22, Okuma:ARCMP23, Gohsrich:E25}. It has been studied across diverse platforms, from photonic systems \cite{Mandal:PRL20, Wang:AOPA23} to electronic devices \cite{Kokhanchik:PRB23, Ochkan:NP24}, and most recently in chiral molecular systems coupled to ferromagnetic leads, where it governs charge trapping and magnetotransport anomalies \cite{Zhao:NC25}.
However, the connection between non-Hermitian eigenvectors and standard transport theory is not trivial. This poses a fundamental challenge to understanding how the NHSE manifests in electronic transport \cite{Ghaemi-Dizicheh:PRA21}. To the date, it has been shown that it induces persistent currents \cite{Zhang:PRL20, Yan:PRB24, Shen:PRL24, Pino:PRB25, Yang:25}, and nonreciprocal charge transport in heterojunctions with reservoir-engineered non-Hermicity \cite{Geng:PRB23}.

In this paper, we demonstrate that the NHSE can be experimentally detected in a realistic device based on a dissipative Rashba nanowire. While similar devices have been studied previously in the context of topological superconductivity \cite{San-Jose:SR16, Avila:CP19, Escribano:PRB21, Escribano:nQM22, Arouca:PRB23, Singh:CP23, Cayao:PRB24c,Ghosh:SP24, Gogoi:25}, our model introduces dissipation through coupling to a ferromagnetic reservoir along the wire's length. This approach has been explored for two-dimensional systems in Ref. \cite{Cayao:JPCM23}, and generically for one-dimensional models and large magnetic fields in Ref. \cite{Kokhanchik:PRB23}. 
We address here an intermediate regime and, importantly,  propose a concrete experimental realization for the observation of NHSE through transport spectroscopy.
We calculate the local and nonlocal conductance and show that the former is symmetric, i.e., it is the same at both nanowire ends, while the latter exhibits pronounced nonreciprocity, i.e., it depends on which end serves as the source. We establish that this nonreciprocal behavior is directly linked to the NHSE, demonstrating an explicit connection between transport asymmetry and eigenstate localization phenomena as predicted in Ref. \cite{Geng:PRB23}. We are also able to explain through perturbation theory the difference between the EPs  position in parameter space of the bulk spectrum and that observed in finite-length wires.

\begin{figure}
   \centering
   \includegraphics[width=\columnwidth]{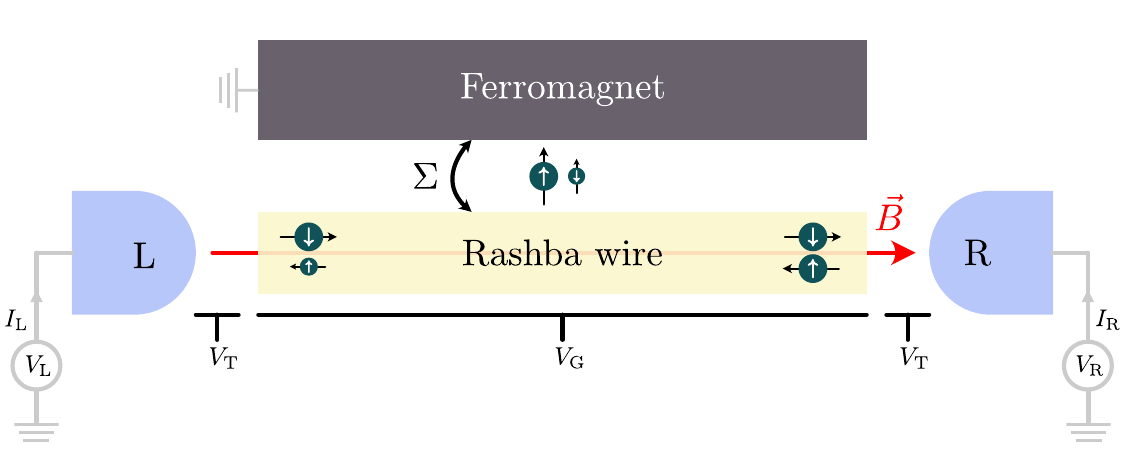}
   \caption{\textbf{Rashba nanowire coupled to a ferromagnetic reservoir transport setup.} Sketch of a semiconductor nanowire of length $L$ (yellow) strongly coupled to a grounded ferromagnet (gray), whose degrees of freedom are included through a spin-dependent self-energy $\Sigma$. The device is contacted by normal metallic leads $\alpha = \left\{\rm L, R\right\}$ (blue) at both sides that allow local and nonlocal transport measurements by varying the applied voltages $V_\alpha$ and measuring the resulting currents $I_\alpha$. Gate voltages $V_T$ and $V_G$ control the coupling strength to the leads and the chemical potential, respectively. An axial magnetic field $\vec{B}$ is applied parallel to the nanowire axis. In the helical regime, left-movers' spin is mostly parallel to the ferromagnet polarization axis ($\uparrow_y$), while right-movers' spin ($\downarrow_y$) is mostly antiparallel. Consequently, dissipation of left movers is favored over right movers. This is depicted schematically over the sketch.
   }
   \label{fig:sketch}
\end{figure}

\cpar{Results}
We consider a semiconductor nanowire with strong Rashba spin-orbit coupling (SOC) that is coupled to a ferromagnetic contact and subjected to an axial magnetic field, see Fig. \ref{fig:sketch}. The device geometry is configured such that the ferromagnet spin polarization aligns parallel to the SOC field and perpendicular to the applied magnetic field \footnote{A small misalignment of the magnetic and SOC fields was also tested, and does not have a qualitative effect on the results.}. While this configuration is unusual, it has been engineered in other contexts \cite{Lee:JAP15, Ai:NC21, other}. Under these conditions,the wire single-particle Hamiltonian takes the form
\begin{equation}
    H_0 = \frac{p_x^2}{2 m^*} - \mu + \alpha p_x \sigma_y + B \sigma_x,
    \label{eq:ham}
\end{equation}
where $\alpha$ is the SOC strength, $\mu$ the electrostatically tunable chemical potential controlled via gate voltage $V_{\rm G}$, $B$ the Zeeman energy that arises from the axial magnetic field $\vec{B}$, $\sigma_i$ the Pauli matrices, and $m^*$ the effective mass, with $\hbar = 1$ and the wire oriented along the $x$ axis. To incorporate the effects of the ferromagnetic contact, we employ the Green’s function formalism and integrate out the ferromagnet degrees of freedom in the wide-band limit  \cite{Datta:95}, yielding the wire retarded Green’s function 
\begin{equation}
    G^R = \left(\omega - H_0 - \Sigma^R(\omega = 0)\right)^{-1} = \left(\omega - H_{\rm eff}\right)^{-1},
\end{equation}
where the ferromagnet physics is captured through the self-energy $\Sigma^R = \mathcal{V}^\dagger g_F^R\mathcal{V}$, with $g_F^R$ the ferromagnet's retarded Green's function and $\mathcal{V}$ the spin-independent coupling to the Rashba wire. In the spin basis aligned with the ferromagnet's polarization axis, $g_F^R$ is diagonal \cite{Bergholtz:PRR19,Cayao:PRB22}, so we model its self-energy as
\begin{equation}
    \Sigma^R(\omega = 0) = -i \left(\gamma_0 +  \gamma_y \sigma_y\right),
\end{equation}
where $\gamma_y$ ($\gamma_0$) is a spin-dependent (independent) dissipation. These quantities must satisfy $\gamma_0 \geq \gamma_y \geq 0$ to preserve causality \footnote{The anti-Hermitian part of $\Sigma_{R}$ has to be negative definite so that the eigenvalues of $H_{\rm eff}$ lie all in the lower half complex plane. This represents decay into the reservoir.}. This self-energy represents the simplest possible dissipative coupling to a featureless ferromagnet. We define $H_{\rm eff} = H_0 + \Sigma^R(\omega = 0)$ as a non-Hermitian effective Hamiltonian.

\begin{figure}
   \centering
   \includegraphics[width=\columnwidth]{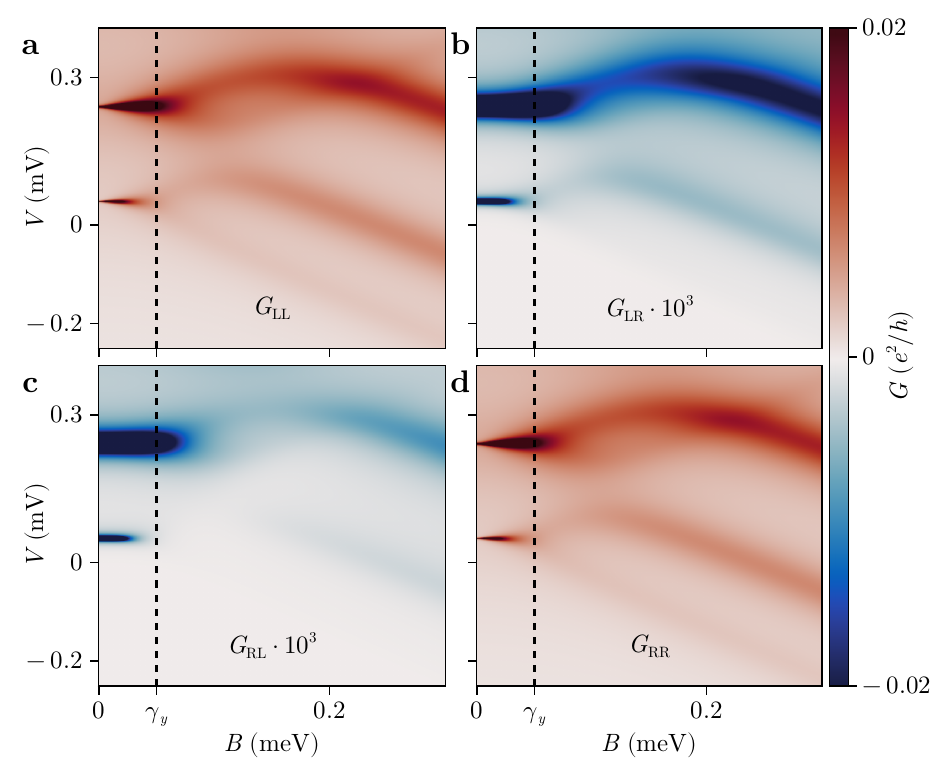}
   \caption{\textbf{Local and nonlocal conductance.} (a, d) Local conductance on both sides of the wire, $G_{\rm{LL}}$ and $G_{\rm{RR}}$, versus source voltage $V$ and Zeeman energy $B$. $G_{\rm{LL}}$ and $G_{\rm{RR}}$ are symmetric (equal). Conductance peaks at low $B$ bifurcate close to $B = \gamma_y$.  $\gamma_y$ quantifies the spin-oriented dissipation. (b, c) Same as (a, d) but for nonlocal conductance. $G_{\rm{LR}}$ and $G_{\rm{RL}}$ are different (nonreciprocal) for large $B$.
   Parameters: $m^* = 0.023 m_e$, $\alpha = 10$meVnm, $\mu = 0$~meV, $L = 500$~nm, $\gamma_0 = \gamma_y = 0.05$~meV, and $a_0=5$~nm.
   }
   \label{fig:cond}
\end{figure}

To simulate a transport experiment, the wire is weakly coupled at both ends to normal leads $\rm L$ and $\rm R$ through gate-tunable barriers $V_{\rm T}$. The conductance matrix, $G_{\alpha\beta} = dI_\alpha /dV_\beta$, with $\alpha, \beta \in \left\{ \rm L, R\right\}$, is calculated using a tight-binding model with discretization parameter $a_0$, combined with the Green's functions formalism. Figure \ref{fig:cond} presents an example of the conductance matrix elements versus Zeeman energy. In this case, we consider transparent barriers between the wire and the contacts, but qualitatively similar results are obtained for other transparencies (not shown).  All four matrix elements display a similar structure, featuring non-dispersive peaks that split into two dispersive ones near $B  =\gamma_y$. We show below that this bifurcation is a topological phase transition at an EP \cite{Solow:25}. The local conductance [panels (a), (d)] is symmetric across all Zeeman energies, $G_{\rm LL} = G_{\rm RR}$, as expected from the symmetry of the device. In contrast, the most striking feature of these calculations is the pronounced nonreciprocity in nonlocal conductance [panels (b), (c)] with $G_{\rm LR} \gg G_{\rm RL}$ for large $B$. This nonreciprocity is not a diode-like effect, rather, it is a measure of different currrent loss for left-moving and right-moving current, and is thus an inherently non-Hermitian effect.

\begin{figure}
   \centering
   \includegraphics[width=\columnwidth]{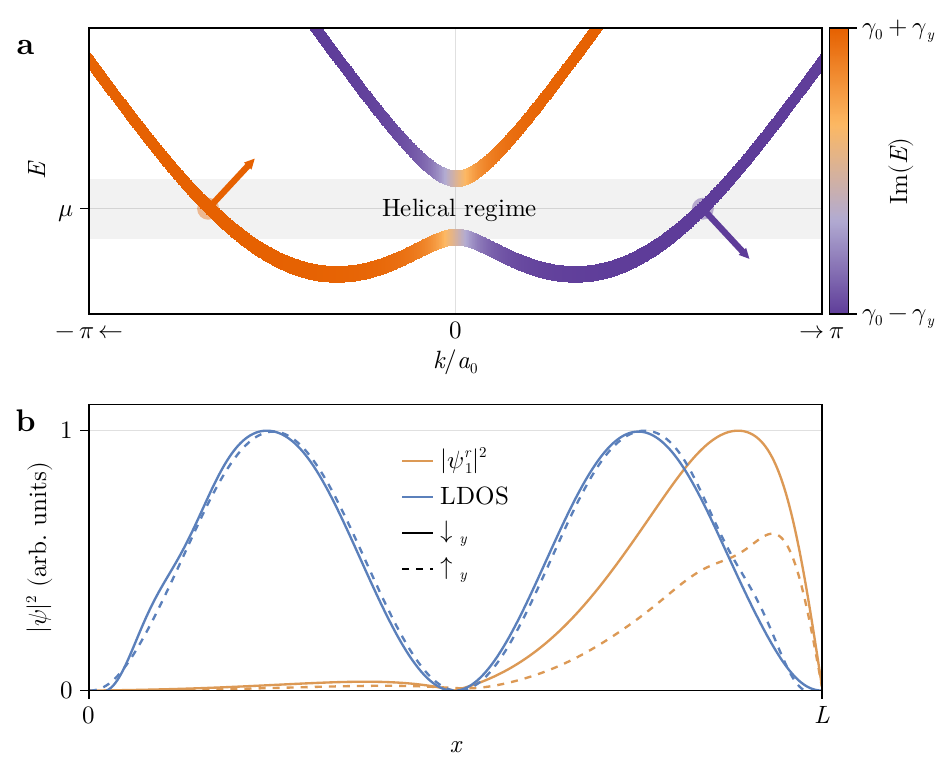}
   \caption{\textbf{Relation of the non-Hermitian skin effect with dissipation.} (a) Real part of the bulk spectrum versus momentum along the wire, colored with the magnitude of its imaginary part. In the helical regime, the spin of the left (right) movers (represented by arrows) is locked to a direction mostly parallel (antiparallel) to the ferromagnet polarization axis $\hat{y}$. Thus, left movers are dissipated more than right movers. (b) Squared norm of the lowest-energy right eigenstates (orange) and LDOS (blue) along the wire axis of the finite-length device. Dashed (solid) lines  distinguish parallel (antiparallel) spin components along the ferromagnet polarization axis. The right eigenstates are pushed to one end of the wire, a manifestation of the NHSE, which indicates the direction favored for transport. The LDOS remains symmetric, thus the local conductance results. Parameters as in Fig. \ref{fig:cond}, except for $L \rightarrow \infty$ in panel (a) and $L=5$$\rm \mu$m in panel (b).
   }
   \label{fig:bands}
\end{figure}

To elucidate the origin of this nonreciprocal conductance, we analyze the complex spectrum of the bulk system, as shown in Fig. \ref{fig:bands}(a). The real part of the spectrum retains the characteristic band structure versus axial momentum $k$ of a Rashba nanowire. However, the self-energy introduces an imaginary component, or dissipation, to the eigenvalues (plotted with color over their real part). This dissipation is spin dependent, resulting in a stronger effect for one spin orientation compared to the other. In the helical regime ($\mu$ within the $k=0$ gap, $B > \gamma_y$), spin and momentum are locked: right movers predominantly possess $\downarrow_y$ spin, while left movers are mainly $\uparrow_y$ (note that there is a spin canting along $\hat{x}$ due to the applied magnetic field). As a result, right movers experience less dissipation than left movers. In turn, this explains the nonreciprocity observed in Fig. \ref{fig:cond}, where $G_{\rm LR} \gg G_{\rm RL}$ for large $B$. Conversely, outside the helical regime ($\mu$ outside the gap or $B < \gamma_y$), spin-momentum locking is absent, and both right and left movers are dissipated equally.

The NHSE also provides a framework for interpreting conductance. Because the effective Hamiltonian is non-Hermitian, its left and right eigenvectors differ \cite{Brody:JPAMT13}. They are defined as $H\ket{\psi_n^r}=E_n\ket{\psi_n^r}$ and $\bra{\psi_n^l}H=\bra{\psi_n^l}E_n$, where $l/r$ is left/right and $E_n$ the eigenvalues. They are moreover taken to satisfy the biorthogonality condition $\braket{\psi^l_n}{\psi^r_m}=\delta_{nm}$. For a finite-length wire, in Fig. \ref{fig:bands}(b) we plot $|\psi_1^r|^2$ (orange lines) and the local density of states $\text{LDOS} = \psi_1^{l\dagger} \psi_1^r$ (blue lines) along the axial direction. In terms of biorthogonal eigenstates, the retarded Green's function can be written as
\begin{equation}
    G^{R}(\omega)=\sum_n\frac{\ket{\psi^r_n}\bra{\psi^l_n}}{\omega-E_n}.
\end{equation}
By expressing the conductance in terms of the Green's function using standard methods (see \ref{Conductance}), one finds that the local and nonlocal conductance take different forms. Assuming weak coupling between the wire and the contacts, the dominant term in the local conductance is
\begin{equation}
    G_{\alpha \alpha}\sim \text{Im}\sum_n\frac{\bra{\psi^l_n}x_\alpha\ket{\psi^r_n}}{eV-E_n},
    \label{eq:local_cond}
\end{equation}
where $x_\alpha$ is the projector at the wire site closest to contact $\alpha$ and $V$ the applied voltage. This expression is proportional to the LDOS close to the contact, just as happens with Hermitian Hamiltonians. Since the LDOS along the wire is symmetric, as seen in Fig. \ref{fig:bands}(b) (note that $\text{LDOS} = \psi^{l \dagger}\psi^r \neq \left| \psi^r \right|^2$), it yields $G_{\rm LL}=G_{\rm RR}$. In contrast, the nonlocal conductance takes the form
    \begin{equation}
        G_{\alpha \beta\neq \alpha}\sim \text{Re} \sum_{mn}\frac{\bra{\psi^l_n}x_\beta\ket{\psi_m^l}\bra{\psi_m^r}x_\alpha\ket{\psi^r_n}}{(eV-E_n)(eV-E_m^*)}.
        \label{eq:nonlocal_cond}
    \end{equation}
To understand this expression, note that the projection operators act on either right or left eigenstates, unlike Eq. \eqref{eq:local_cond} where the projection acts on both. Orange lines in Fig. \ref{fig:bands}(b) show that right eigenstates accumulate at one system boundary ($\rm R$ contact in this case), making the projection large on that edge and negligible on the opposite side. Conversely, left eigenstates localize on the opposite boundary for the same eigenvalue (not shown),  guaranteed by biorthogonality \cite{Brody:JPAMT13}. Hence, nonlocal conductance is maximized when $\alpha$ and $\beta$ lie on the boundaries favored by right and left eigenstates, respectively, and minimized in the reverse configuration \footnote{Notice that Fig. \ref{fig:bands}(b) is calculated for a wire decoupled from $\rm L, R$ contacts, hence all norms vanish at the wire ends. However, when the coupling to the contacts is finite, these quantities are non-zero.}. This contrasts with the local conductance, where the biorthogonal projection in Eq. \eqref{eq:local_cond} cancels out the left/right localization \cite{Brody:JPAMT13}. This asymmetry is absent in Hermitian systems, where left and right eigenstates are identical.

\begin{figure}
   \centering
   \includegraphics[width=\columnwidth]{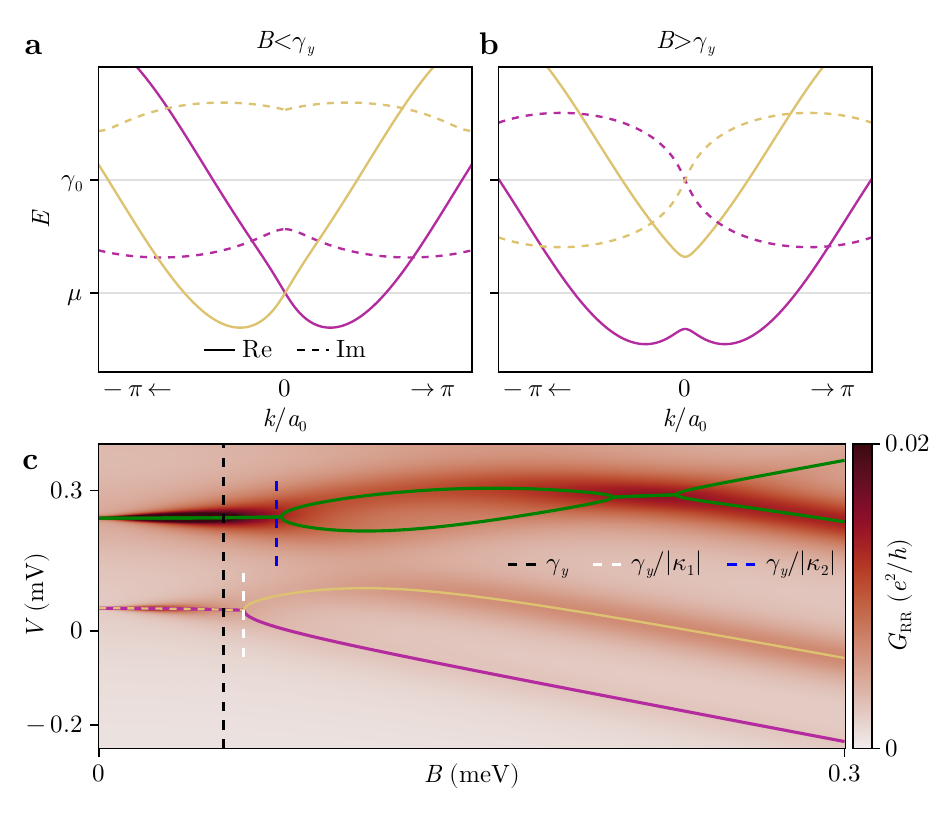}
   \caption{\textbf{Non-Hermitian topological phase transition.} (a, b) Spectrum of the bulk system before (a) and after (b) the topological phase transition at $B = \gamma_y$. In (a), there is a gap in the imaginary part around the spin-independent dissipation $\gamma_0$, while in (b), there is a gap in the real part around the chemical potential $\mu$. (c) Local conductance at one end of the device versus Zeeman energy $B$. The four lowest-energy levels are overplotted with different colors. Vertical black dashed line indicates the bulk topological phase transition. For the lower (upper) pair, Eq. \eqref{eq:kappa} predicts an exceptional point at the white (blue) dashed line, matching the conductance peak bifurcations. Parameters as in Fig. \ref{fig:cond} except for $L \rightarrow \infty$ for panels (a, b)
   }
   \label{fig:trans}
\end{figure}
The non-Hermitian Rashba wire belongs to the symmetry class AI \cite{Kokhanchik:PRB23}, where systems with a point gap exhibit a $\mathbb{Z}$ topological invariant. It is given by the winding number of the complex spectrum around the point gap. For two-band models with PBC this winding number reduces to a $\mathbb{Z}_2$ invariant, which can be expressed as \cite{Budich:PRB19}
\begin{equation}
    w_{\rm PBC}=\mathrm{sign}\det(\Vec{d}(0)\cdot\Vec{\sigma})=\mathrm{sign}(B^2-\gamma_y^2),
\end{equation}
with $H_{\rm eff} = d_0(k) +\vec{d}(k)\cdot \vec{\sigma}$. This is well defined except at the EP $\abs{B}~=~\abs{\gamma_y}$. Figure \ref{fig:trans}(a, b) illustrates this topological phase transition in the spectrum. Panel (a) shows the imaginary-gap phase ($B < \gamma_y$), while panel (b) shows the real-gap phase ($B > \gamma_y$).

In Fig. \ref{fig:trans}(c), we overlay the real part of the finite wire’s energy spectrum, i.e. with OBC, onto the conductance plot of Fig. \ref{fig:cond}(a). Each conductance peak corresponds to the real part of an eigenvalue of the effective Hamiltonian. At small Zeeman energies, each peak corresponds to two eigenvalues with the same real part but different imaginary part. As $B$ increases, each pair of eigenvalues coalesces at an EP, closing the imaginary  gap and opening a real one. Although this represents a non-Hermitian topological phase transition analogous to the PBC case, the EP for each pair occurs at different Zeeman fields and always for $B>\gamma_y$. Such level-dependent shifts have been observed in other systems  \cite{Zeng:PRA16, Yao:PRL18, Li:JPCM22}, but they have not been explained so far. 

To elucidate these shifts, we consider the OBC system at $B=0$, where one can obtain closed-form eigenstates (see \ref{ex-point-shift}). Within the two-state subspace that coalesces at an EP, we treat $B$ as a perturbation. To first order in $B$, the effective Hamiltonian reads
\begin{equation}
    \tilde{H}^{(1)}_{\rm eff}=\begin{pmatrix}
        E_0+i\gamma_y & \kappa B \\\kappa^* B & E_0-i\gamma_y
    \end{pmatrix},
    \label{eq:perturbed}
\end{equation}
where $\kappa=\bra{\psi^+}\sigma_x\ket{\psi^-}$ is the overlap between the unperturbed states. Calculating the topological invariant as above, we find that the invariant for the $\nu$th pair is given by
\begin{equation}
    w_{\rm OBC}^{\nu}=\mathrm{sign}(B^2\abs{\kappa_\nu}^2-\gamma_y^2),
\end{equation}
with
\begin{equation}
    \abs{\kappa_\nu}^2=\frac{\sin(\alpha mL)}{\alpha mL\left(1-\frac{\left(\alpha mL\right)^2}{\nu^2\pi^2}\right)}.
    \label{eq:kappa}
\end{equation}
The vertical dashed lines in Fig. \ref{fig:trans}(a) mark the EP positions predicted by Eq. \eqref{eq:kappa}, demonstrating excellent agreement between the analytical approximation and the exact numerical calculations.

\cpar{Conclusions}
We have proposed an experimental setup capable of detecting the non-Hermitian skin effect (NHSE) through transport measurements. It consists of a Rashba nanowire strongly coupled to a ferromagnetic lead with polarization axis parallel to the wire's SOC field. We have found that this system exhibits symmetric local conductance, but nonreciprocal nonlocal conductance after a non-Hermitian topological phase transition. This can be interpreted either as the accumulation of non-Hermitian eigenstates at one edge of a quantum wire (NHSE) or, equivalently, as a manifestation of directional dissipation mediated by spin. Additionally, we have shown that the Zeeman energy at which EPs happen with OBC is shifted relative to the bulk system with PBC. This shift can be accurately anticipated via perturbative analysis in the $B$-field. Our findings demonstrate a connection between the nonlocal electrical conductance and the NHSE.  We have also shown that the non-Hermitian physics perspective can unveil novel features of these systems, such as nonreciprocal conductance, that may remain concealed in traditional Hermitian descriptions.

Lastly, we remark that the NHSE is a topological feature dependent on the presence of a point gap in the complex eigenvalue spectrum \cite{Gohsrich:E25}. Therefore, it persists under any disorder that does not close said gap.

\cpar{Acknowledgments}
We thank Pablo San-Jose for insightful discussions. This research was supported by Grants No. PID2021-125343NB-I00, PRE2022-101362 and CEX2024-001445-S, funded by MICIU/AEI/10.13039/501100011033, ``ERDF A way of making Europe'' and ``ESF+''; the European Research Council (Grant Agreement No. 856526) and by the DFG Collaborative Research Center 183, Project No. 277101999.

\cpar{Data availability}
The code to perform the calculations is available in Ref.  \cite{Paya:25b}. Transport calculations make use of Quantica.jl  \cite{San-Jose:25}. Visualizations were made with the Makie.jl package  \cite{Danisch:JOSS21}. 

\appendix

\appendixpar{Conductance in non-Hermitian systems}
\label{Conductance}
To calculate the conductance in a non-Hermitian system, we start with the standard expression for the current through lead $\alpha$ \cite{Haug:07}
\begin{equation}
    \langle I_\alpha \rangle=-e\Re{\int \frac{d\omega}{2\pi}\Tr[G^R(\omega)\Sigma_\alpha^<(\omega)+G^<(\omega)\Sigma_\alpha^A(\omega)]}
\end{equation}
where $G^{R/<}(\omega)$ is the retarded/lesser Green's function of the system and $\Sigma^{A/<}_\alpha(\omega)$ is the advanced/lesser tunneling self-energy of lead $\alpha$. We consider leads in the wide-band approximation, so $\Sigma_\alpha^A=-i\Gamma_\alpha x_\alpha$ where $\Gamma_\alpha$ is the coupling to lead $\alpha$ and $x_\alpha$ is a projection operator which projects onto the point where lead $\alpha$ couples to the system. Using $G^<(\omega)=G^R(\omega)\Sigma^<(\omega)G^A(\omega)$ valid for non-interacting systems, one obtains the differential conductance
\begin{widetext}
\begin{equation}
    G_{\alpha \beta}=\frac{d\langle I_\alpha \rangle}{dV_\beta}=-2e^2 \text{Re}\int \frac{d\omega}{2\pi}f'(\omega-eV_\beta)\text{Tr}\left[G^R(\omega)(i\Gamma_\alpha x_\alpha)\delta_{\alpha \beta}+G^R(\omega)(\Gamma_\beta x_\beta)G^A(\omega)\Gamma_\alpha x_\alpha \right]
\end{equation}
\end{widetext}
where $f(\omega)$ is the Fermi function.
In the biorthogonal basis of left and right eigenvectors and taking the zero-temperature limit, with $f'(x)=\delta(x)$, the conductance matrix is
\begin{equation}
    \begin{split}
         G_{\alpha \beta} =& -\frac{e^2}{\pi} \left[ \text{Im} \sum_n\frac{\Gamma_\alpha \bra{\psi_n^l} x_\alpha \ket{\psi_n^r}}{(eV-E_n)}\delta_{\alpha \beta} + \right. \\
         & \left. \text{Re} \sum_{nm}\frac{\Gamma_\beta\Gamma_\alpha\bra{\psi_n^l} x_\beta \ket{\psi_m^l}\bra{\psi_m^r} x_\alpha\ket{\psi_n^r}}{(eV-E_n)(eV-E^*_m)} \right].
    \end{split}
\end{equation}
Physically, the first term is proportional to the LDOS near contact $\alpha$, so the local conductance is primarily a probe of the LDOS, as expected from the Hermitian case. The second term is a measure of whether electrons entering the system from contact $\beta$ can leave the system through contact $\alpha$. To do this, there must be states near contact $\beta$ which time evolve into states near contact $\alpha$. Since $G^{R}$ maps kets to right eigenstates of the Hamiltonian, it is the right eigenstates which determine the forwards time evolution of the system. As such, transport is large towards the side of the system where the right eigenstates are localized.
Considering the expression for the current in a three-leads system, where one lead is integrated out providing the non-Hermitian term, we observe that $G_{\alpha \beta}\neq G_{\beta \alpha}$ implies that $\ket{\psi^r_n}\neq\ket{\psi^l_n}$ for at least some state $n$. In other words, the nonreciprocity of the nonlocal conductance implies the NHSE. 
\\
\appendixpar{Exceptional point shift}
\label{ex-point-shift}
To understand the shifts with Zeeman energy of EPs at OBC systems with respect to the PBC prediction, we treat $B$ as a perturbation within the two-state subspace taht coalesces at each EP.
We begin with the unperturbed Hamiltonian at $B = 0$,
\begin{equation}
    H_0=\begin{pmatrix}
        \frac{p_x^2}{2m^*} + i\gamma_0 & -i\alpha p_x+\gamma_y \\i\alpha p_x-\gamma_y & \frac{p_x^2}{2m^*} + i\gamma_0
    \end{pmatrix},
\end{equation}
with parameters as in Eq. \eqref{eq:ham}. Solving for OBC, $\psi(x = 0) = \psi(x = L) = 0$, we find the eigenvalues,
\begin{equation}
    E_\nu^{\pm} = E_0 \pm i \gamma_y,
\end{equation}
where
\begin{equation}
    E_0=\frac{\pi^2\nu^2}{2m^*L^2}-\frac{\alpha^2 m^*}{2},
\end{equation}
and $\nu$ is a pair index, with eigenvectors
\begin{equation}
    \psi_\nu^{\pm}(x)=\frac{1}{\sqrt{L}}\sin\left(\frac{\nu\pi}{L}x\right)e^{\pm im^*\alpha x}\begin{pmatrix}
        1 \\ \mp i
    \end{pmatrix}.
\end{equation}
The two states $\psi_\nu^\pm$ coalesce at an EP at $\gamma_y = 0$, forming the subspace for our perturbation analysis. The first order in $B$ effective Hamiltonian becomes Eq. \eqref{eq:perturbed}, where the overlap between unperturbed states, $\kappa_\nu$ is given by
\begin{equation}
    \kappa_\nu=\frac{2}{L}\int_0^L dx\sin^2\left(\frac{\nu\pi}{L}x\right)e^{2im\alpha x}.
\end{equation}
Solving this integral analytically yields Eq. \eqref{eq:kappa}.

\bibliography{Non-Hermicity, NHSE_transport_Rashba_nanowire}

\end{document}